\documentclass[paper,12pt]{article}
\begin{document}
\topmargin 0pt \oddsidemargin 0mm

\renewcommand{\thefootnote}{\fnsymbol{footnote}}
\begin{titlepage}

\vspace{2mm}
\begin{center}
{\Large \bf Probing the Vacuum Structure of Spacetime}
 \vspace{12mm}

{\large Sang Pyo Kim\footnote{e-mail address: sangkim@kunsan.ac.kr}}\\

\vspace{5mm}
 {\em Department of Physics, Kunsan National University, Kunsan 573-701,
 Korea\footnote{Permanent Address}}\\
 {\em Institute of Astrophysics, Center for Theoretical Physics,\\Department of Physics, National Taiwan University,
Taipei 106, Taiwan}
\date{\today}
\vspace{5mm}\\
{(Dated: January 21, 2011)}

\end{center}

\vspace{10mm} \centerline{{\bf{Abstract}}} \vspace{5mm}
We explore the question of how to probe the vacuum structure of space time by a massive scalar field through interaction with background gravitons.
 Using the $\Gamma$-regularization for the in-/out-state formalism, we find the effective action of a scalar field in a conformally, asymptotically flat spacetime and a four-dimensional de Sitter space, which is a gravitational analog of the Heisenberg-Euler and Schwinger effective action for a charged scalar in a constant electric field. The effective action is nonperturbative in that it sums all one-loop diagrams with arbitrary number of external lines of gravitons. The massive scalar field becomes unstable due to particle production, the effective action has an imaginary part that determines the decay rate of the vacuum, and the out-vacuum is unitarily inequivalent to the in-vacuum.
\end{titlepage}

\newpage
\renewcommand{\thefootnote}{\arabic{footnote}}
\setcounter{footnote}{0} \setcounter{page}{2}

\section{Introduction}

One of the longstanding problems in theoretical physics is to understand what the vacuum is and how it interacts with matter fields. To probe the vacuum of the Minskowski spacetime, that is, the Dirac sea of particle-antiparticle pairs, one has to turn on a strong (gauge) field and produce pairs of particle and antiparticle interacting with it. The vacuum structure of a curved spacetime, on which laws of physics are formulated, becomes more complicated and intangible than the Minkowski one. Without quantum gravity at present, an effective field theory for gravity may be employed, where matter fields are quantized on the background spacetime.

In this paper, we shall explore by a massive scalar field the vacuum structure of spacetime, in particular, a conformally, asymptotically flat spacetime and a de Sitter (dS) spacetime in four dimensions. As a few leading Feynman diagrams do not exhibit the vacuum structure, we need a nonperturbative method for the effective action in analogy with quantum electrodynamics (QED). Heisenberg-Euler and Schwinger effective action, for instance, probes the vacuum of Dirac sea of particle and antiparticle pairs by a strong external electromagnetic field \cite{Heisenberg-Euler,Schwinger}. The Dirac sea will be explored by strong light sources in the near future \cite{Ringwald}.

To find the nonperturbative effective action for gravity,
 we shall work in the in-/out-state formalism from Schwinger's variational principle. The variational principle states that either interactions or curved spacetimes may make the out-vacuum different from the in-vacuum and the scattering matrix between the out- and the in-vacua provides the effective action \cite{Schwinger-var}. Nikishov and DeWitt showed that the $S$-matrix could be entirely determined by the coefficient for the Bogoliubov transformation between the in- and the out-vacua \cite{Nikishov,DeWitt}. In other formulation one has to integrate the Feynman propagator in the DeWitt-Schwinger representation with respect to mass squared to obtain the effective action (for review and references, see Ref. \cite{Birrel-Davies}).

Recently, Kim, Lee and Yoon have further developed the in-/out-state formalism through the gamma function- or the $\Gamma$-regularization and found the QED effective actions in strong electromagnetic fields that act for a finite period \cite{KLY08}. (For other attempts using the gamma function, see Refs. \cite{AHN,Nikishov03}.) The $\Gamma$-regularization has been elaborated to spatially localized electromagnetic fields \cite{KLY10}, which require the tunneling boundary condition on the solution \cite{Nikishov}. The new formalism has also been applied to dS space minimally coupled to a massive scalar field and yielded the effective action in any dimensional dS \cite{Kim10}, which is the gravitational analog of the Heisenberg-Euler/Schwinger effective action in QED for a constant electric field.

The reason for studying the vacuum polarization (the real part of the effective action) of a conformally, asymptotically flat and a dS space is that firstly, these spacetimes allow one to explicitly find the Bogoliubov transformation and secondly, they suffer from instability due to cosmic particle production, the so-called  ``cosmic laser'' \cite{Polyakov}.
The main difference from the earlier work, Ref. \cite{Kim10}, is that we analyze the divergent and the convergent structures of the effective action for gravity in four dimensions only and interpret the results from the view point of cosmology. The divergent terms are closely related with renormalization of the vacuum energy, the gravitational constant and the one-loop coupling constants.

The organization of this paper is as follows. In Section 2, we formulate the $\Gamma$-regularization for effective actions. In Section 3, we regularize and investigate the UV and the IR structures of Heisenberg-Euler/Schwinger effective action in scalar QED. We study the effective action for a conformally, asymptotically flat spacetime in Section 4 and for a four dimensional dS space in Section 5 and interpret the divergent terms in terms of renormalization of the cosmological constant, the gravitational constant and the coupling constants of one-loop.

\section{$\Gamma$-Regularization for the In-/Out-State Formalism}

The in-vacuum under the influence of an external field, such as a gauge field or a background spacetime, evolves to another one, which needs not necessarily the same as the in-vacuum, and settles in the out-vacuum when the external field acts for a finite period of time. This is true not only for QED in electromagnetic fields and but also for curved spacetimes, in particular, expanding spacetimes and black holes. Further, if the external field produces particles (pairs), the in-vacuum suffers from instability and decays as (the upper/lower sign for bosons/fermions) [in units of $c = \hbar = 1$]
\begin{eqnarray}
|\langle 0, {\rm out} \vert 0, {\rm in} \rangle|^2 = \exp \Bigl[\mp {\cal V} {\cal T} \sum_{\bf K} (1 \pm {\cal N}_{\bf K}) \Bigr], \label{decay}
\end{eqnarray}
where ${\cal V}$ is the space volume, ${\cal T}$ is the duration of the interaction,
${\bf K}$ denotes all quantum states, and ${\cal N}_{\bf K}$ is the number of particles (pairs) produced for ${\bf K}$. In the limit of infinite volume or time, the out-vacuum is orthogonal to and unitarily inequivalent to the in-vacuum. Thus, we need the in-/out-state formalism for the effective action in background external fields.

The effective action, in principle, can be obtained from the generating functional in a curved spacetime
\begin{eqnarray}
Z[J] = \int {\cal D}[\phi] \exp\Bigl[i S_{\rm m} [\phi] + i\int \sqrt{-g} d^4 x J \phi \Bigr],
\end{eqnarray}
where $\phi$ denotes a boson or fermion field, $S_{\rm m} [\phi]$ is the action for the field and $J$ is a source.
The variational principle by Schwinger prescribes \cite{Schwinger-var}
\begin{eqnarray}
\delta Z[0] = i \int {\cal D}[\phi] \delta S_{\rm m} [\phi] e^{iS_{\rm m} [\phi]} = i \langle 0, {\rm out} \vert \delta S_{\rm m} \vert  0, {\rm in} \rangle,
\end{eqnarray}
and thereby the effective action as the $S$-matrix
\begin{eqnarray}
e^{iW} = \langle 0, {\rm out} \vert  0, {\rm in} \rangle = Z[0]. \label{eff-def}
\end{eqnarray}
The energy-momentum tensor follows from varying $W$ with respect to $g^{\mu \nu}$. To compute the effective action (\ref{eff-def}), one needs some method either for the path integral including particles production or for the out-vacuum.

Schwinger himself evaluated the path integral for a charged scalar or fermion in a constant electromagnetic field and found the so-called Heisenberg-Euler/Schwinger effective action in the proper-time integral \cite{Schwinger}. One prominent feature of the effective action is the imaginary part that determines the decay rate, as shown in eq. (\ref{decay}). The other interesting property is the vacuum polarization, where the Minkowski vacuum becomes a medium since the Dirac sea produces and annihilates the (virtual) pairs of particle and antiparticle. The effective action beyond constant electromagnetic fields has been an issue of extensive study in theoretical physics \cite{Dunne-rev}, which belongs to strong field physics, and where nonperturbative quantum effects are important.

The effective action for a massive field can also be found from the Feynman propagator in the DeWitt-Schwinger representation as \cite{Birrel-Davies}
\begin{eqnarray}
W = \frac{i}{2} \int_{m^2}^{\infty} dm^2 \int \sqrt{-g} d^4x G^{\rm DS}_{\rm F} (x,x). \label{feyn}
\end{eqnarray}
The resolvent method for the effective action in QED in a constant and a pulselike
electric field \cite{Dunne-Hall99} and the Green function method for the Coulomb gauge \cite{Nikishov} is
a variation of this method. Nikishov and DeWitt showed that the effective action could be expressed
in terms of the Bogoliubov coefficients (the upper/lower sign for bosons/fermions) as \cite{Nikishov,DeWitt}
\begin{eqnarray}
W = \int \sqrt{-g} d^4 x {\cal L}_{\rm eff} = - i \ln (\langle 0, {\rm out} \vert 0, {\rm in}
\rangle) = \pm i ({\cal V} {\cal T}) \sum_{\bf K} \ln (\mu_{\bf K}^*), \label{bog-eff}
\end{eqnarray}
since the out-vacuum is related to
the in-vacuum through the Bogoliubov transformation for the bosons/fermions interacting the external field
\begin{eqnarray}
a_{{\rm out}, {\bf K}}= \mu_{\bf K} a_{{\rm in}, {\bf K}} + \nu_{\bf K}^* a_{{\rm in}, {\bf K}}^{\dagger}. \label{bog tran}
\end{eqnarray}
These coefficients satisfy the relation for each ${\bf K}$
\begin{eqnarray}
| \mu_{\bf K} |^2 \pm | \nu_{\bf K}|^2 = 1,
\end{eqnarray}
which follows from the commutation/anticommutation relation for bosons/fermions or
spin-statistics theorem. The effective action (\ref{bog-eff}) even after renormalization constrains the vacuum persistence
(twice of the imaginary part) and the number of produced particles (pairs), ${\cal N}_{\bf K}=
| \nu_{\bf K}|^2$, as \cite{KLY08,KLY10,GGT,Hwang-Kim}
\begin{eqnarray}
2 {\rm Im}({\cal L}_{\rm eff}) =  \pm \sum_{\bf K} \ln (1 \pm  {\cal N}_{\bf K}). \label{gen-rel}
\end{eqnarray}

The key idea of the gamma function- or the $\Gamma$-regularization is that most of known solutions for free fields in background fields can be expressed in the hypergeometric function, whose connection formula in two regions determines the Bogoliubov coefficients as a rational function of gamma functions with complex arguments, and that the logarithm of the gamma function \cite{Gamma},
\begin{eqnarray}
\ln \Gamma (\alpha \pm i \beta) = \int_0^{\infty} \Bigl[\frac{e^{- (\alpha \pm i \beta) z}}{1 - e^{-z}} -
\frac{e^{-z}}{1 - e^{-z}} + (\alpha \pm i \beta -1) e^{-z} \Bigr] \frac{dz}{z},\label{gamma}
\end{eqnarray}
has an analytic continuation in the complex plane and the residue theorem determines the imaginary part.
The second and the third terms are to be regulated away through renormalization. The residue theorem for the first term as a complex function of $z$, assuming $ \beta \geq 0$, leads to
\begin{eqnarray}
\int_0^{\infty} \frac{dz}{z} \frac{e^{- (\alpha \pm i \beta) z}}{1 - e^{-z}} = {\cal P} \int_0^{\infty} \frac{dz}{z} \frac{e^{- (\alpha \pm i \beta) (\mp iz)}}{1 - e^{\pm i z}} \mp \pi i \sum_{n =1}^{\infty}\frac{e^{- (\alpha \pm i \beta) (\mp 2 n \pi i)}}{(\mp) 2 n \pi i}, \label{G-reg}
\end{eqnarray}
where the upper/lower sign is for the contour of a quarterly circle in the fourth/first quadrant and ${\cal P}$ denotes the principal value. Then, with a proper regularization through renormalization, we obtain the effective action in QED and dS spaces.

\section{Regularization of Heisenberg-Euler/Schwinger Effective action}

 In scalar QED a charged massive scalar in a constant electric field in the vector gauge has the Bogoliubov coefficient for the momentum transverse to the electric field \cite{KLY08,KLY10}
\begin{eqnarray}
\mu_{{\bf k}_{\perp}} = \frac{\sqrt{2 \pi} e^{- i \frac{\pi}{4}} e^{- \pi \frac{m^2+ {\bf k}_{\perp}^2}{4 qE}}}{\Gamma ( \frac{1}{2} + i \frac{m^2+ {\bf k}_{\perp}^2}{2qE} )}.
\end{eqnarray}
Neglecting the terms that are to be regulated away through renormalization and using the $\Gamma$-regularization (\ref{G-reg}) in eq. (\ref{bog-eff}), we obtain the effective action 
\begin{eqnarray}
{\cal L}_{\rm eff} (E) = - \frac{qE}{2 (2 \pi)} \int \frac{d^{2}
{\bf k}_{\perp}}{(2 \pi)^{2}} \Bigl[ {\cal P} \int_0^{\infty} \frac{ds}{s}
\frac{e^{- \frac{m^2+ {\bf k}_{\perp}^2}{2qE} s}}{\sin(\frac{s}{2})}
 -i \ln (1 + {\cal N}_{{\bf k}_{\perp}} ) \Bigr], \label{HES eff}
\end{eqnarray}
where the number of pairs via Schwinger mechanism is
\begin{eqnarray}
{\cal N}_{{\bf k}_{\perp}} = \langle 0, {\rm out} \vert a^{\dagger}_{{\rm in}, {\bf k}_{\perp}} a_{{\rm in}, {\bf k}_{\perp}} \vert  0, {\rm out} \rangle = e^{- \pi \frac{m^2+ {\bf k}_{\perp}^2}{qE}}.
\end{eqnarray}
Integrating the momentum integral, we get the Heisenberg-Euler/Schwinger effective action for scalar QED
\begin{eqnarray}
{\cal L}_{\rm eff} (E) = - \frac{1}{2} \Bigl(\frac{qE}{2\pi} \Bigr)^{2} {\cal P}
\int_0^{\infty} \frac{ds}{s^{2}}  \frac{e^{ - \frac{m^2}{2qE} s}
}{\sin(\frac{s}{2})} -
\frac{i}{2 (2\pi)^{3}} \sum_{n=1}^{\infty} \Bigl( \frac{qE}{n} \Bigr)^{2}
(-e^{-\frac{\pi m^2}{qE}})^{n}.
\label{HES eff2}
\end{eqnarray}

To investigate the UV and IR structures of the effective action, we expand ${\rm cosec} (s/2)$ in power series of $s$ and express the vacuum polarization (the real part of the effective action) in the form
\begin{eqnarray}
{\cal L}_{\rm eff} (E) &=& - \Gamma (-2) \frac{m^4}{(4 \pi)^2} - \Gamma (0) |B_{2}| \frac{(qE)^2}{(4\pi)^2}
\nonumber\\&& - \frac{1}{2} \Bigl(\frac{qE}{2\pi} \Bigr)^{2} \sum_{n = 2}^{\infty} \frac{(2^{2n-1} -1) (2n-3)!|B_{2n}|}{2^{2n-1} (2n)!}  \Bigl(\frac{qE}{m^2} \Bigr)^{2n-2},
\end{eqnarray}
where $B_{2n}$ is the Bernoulli number. The first and the second terms renormalize the vacuum energy and the charge for the Maxwell term, respectively, and are removed by subtracting $(2/s - s/12)$ from ${\rm cosec}(s/2)$.
It is interesting to note that the vacuum polarization combined for scalar and spinor QED with the same spin multiplicity $(2 |\sigma|+1)$ regulates away the vacuum energy as in supersymmetry theory
\begin{eqnarray}
{\cal L}_{\rm eff} (E) = - \frac{2|\sigma|+ 1}{2} \Bigl(\frac{qE}{2\pi} \Bigr)^{2} {\cal P}
\int_0^{\infty} \frac{ds}{s^{2}}  e^{ - \frac{m^2}{2qE} s} \frac{1- \cos (\frac{s}{2})}{\sin(\frac{s}{2})}.
\label{HES eff3}
\end{eqnarray}

\section{Conformally, Asymptotically Flat Spacetime}

Conformally flat spacetimes with two asymptotically flat regions provide a good model to study the effective action for gravity since the in- and the out-vacua are Minkowski ones after rescaling time and length and are connected by a one-parameter conformal vacua between the two regions. The conformally flat spacetime under study with the metric \cite{Birrel-Davies}
\begin{eqnarray}
ds^2 = - C(\eta) (d \eta^2 - d {\bf x}^2), \quad C (\eta) = A + B \tanh \Bigl( \frac{\eta}{T} \Bigr), \label{conf}
\end{eqnarray}
has two asymptotically flat regions and allows an explicit solution for a massive minimal scalar field.
Using the freedom of rescaling time, one can set $A =1$ and restrict $B < 1$ from causality. Then,
the remaining two free parameters $B$ and $T$ may be determined by the scalar curvature and the Ricci tensor squared:
\begin{eqnarray}
R = - 3 \frac{\ddot{C}}{C^2} + \frac{3}{2}\frac{\dot{C}^2}{C^3}, \quad R^{\mu \nu} R_{\mu \nu} = \frac{9}{4C^2} \Bigl(\frac{\ddot{C}}{C} - \frac{\dot{C}^2}{C^2} \Bigr)^2 + \frac{3}{4C^2}\Bigl( \frac{\ddot{C}}{C} \Bigr)^2.
\end{eqnarray}
For instance, at $\eta = 0$, the geometric invariants take $R = (3/2) (B/T)^2$ and $R^{\mu \nu} R_{\mu \nu}=R^2$, with $R^{\alpha \beta \mu \nu}R_{\alpha \beta \mu \nu}$ being determined by Gauss-Bonnet theorem.

To determine the Bogoliubov coefficients, we use the solution for a massive minimal scalar \cite{Birrel-Davies}
\begin{eqnarray}
\varphi_{\bf k}(\eta, {\bf x}) &=& \frac{1}{\sqrt{2 \pi \omega_{\rm in}}} e^{- i \omega_+ \eta + i {\bf k} \cdot{\bf x} - i \omega_-T \ln (2 \cosh \frac{\eta}{T})} \nonumber\\&& \times  F(1+ i \omega_-T, i \omega_-T; 1 - i \omega_{\rm in} T; \frac{1}{2} (1 + \tanh \frac{\eta}{T})), \label{m-sol}
\end{eqnarray}
where $F$ is the hypergeometric function and the frequencies at the past and the future infinity and their average and difference are given by
\begin{eqnarray}
\omega_{\rm in/out} ({\bf k})  = \sqrt{{\bf k}^2 + m^2 (1 \mp B)}, \quad
\omega_{\pm} ({\bf k}) = \frac{1}{2} (\omega_{\rm out} \pm \omega_{\rm in}).
\end{eqnarray}
The solution (\ref{m-sol}) describes the incoming positive frequency solution at the past infinity $(\eta = - \infty)$
\begin{eqnarray}
\varphi_{\bf k}(\eta,{\bf x}) &=& \frac{e^{- i \omega_+ \eta + i {\bf k} \cdot {\bf x}}}{\sqrt{2 \pi \omega_{\rm in}}}.
\end{eqnarray}
At the future infinity $(\eta = \infty)$ the solution analytically continues to the following form
\begin{eqnarray}
\varphi_{\bf k} (\eta, {\bf x}) = \frac{e^{- i \omega_{\rm out} \eta + i {\bf k} \cdot {\bf x}}}{\sqrt{2 \pi \omega_{\rm out}}}  \mu_{\bf k}
+ \frac{e^{ i \omega_{\rm out} \eta + i {\bf k} \cdot {\bf x}}}{\sqrt{2 \pi \omega_{\rm out}}}\nu_{\bf k},
\end{eqnarray}
where
\begin{eqnarray}
\mu_{\bf k} = \sqrt{\frac{\omega_{\rm out}}{\omega_{\rm in}}} \frac{\Gamma (1 - i \omega_{\rm in} T)\Gamma (-i \omega_{\rm out} T) }{\Gamma (1 - i \omega_+ T)\Gamma (-i \omega_+ T) }, ~
\nu_{\bf k} = \sqrt{\frac{\omega_{\rm out}}{\omega_{\rm in}}} \frac{\Gamma (1 - i \omega_{\rm in} T)\Gamma (i \omega_{\rm out} T) }{\Gamma (1 + i \omega_- T)\Gamma (i \omega_- T) }.
\end{eqnarray}

Now the effective action per unit volume and per unit time $(W/{\cal VT} = {\cal L}_{\rm eff})$ in the in-/out-state formalism is given by
\begin{eqnarray}
{\cal L}_{\rm eff} &=& i \int \frac{d^3{\bf k}}{(2 \pi)^3} \Bigl[ \ln \Gamma (1 + i \omega_{\rm in} T) + \ln \Gamma (i \omega_{\rm out} T)
\nonumber\\&&
- \Gamma (1 + i \omega_+ T) - \Gamma (i \omega_+ T) + \frac{1}{2} (\ln\omega_{\rm out} - \ln \omega_{\rm in})  \Bigr].
\end{eqnarray}
The last term will be regulated away by renormalization. The $\Gamma$-regularization (\ref{G-reg}) gives the effective action in a complex plane
\begin{eqnarray}
{\cal L}_{\rm eff} = i \int \frac{d^3{\bf k}}{(2 \pi)^3} \int_{0}^{\infty} \frac{dz}{z} \frac{F(z)}{1-e^{-z}},
\end{eqnarray}
where
\begin{eqnarray}
F(z) = e^{-(1 + i \omega_{\rm in} T)z} + e^{-i \omega_{\rm out} T z} - e^{-(1 + i \omega_+ T)z} - e^{-i \omega_+ Tz}.
\end{eqnarray}
Particle production makes the vacuum unstable, so $F(z)$ is complex and thereby the effective action has an imaginary part. According to the $\Gamma$-regularization, we have
\begin{eqnarray}
\int_{0}^{\infty} \frac{dz}{z} \frac{F(z)}{1- e^{-z}} = {\cal P} \int_{0}^{\infty} \frac{ds}{s} \frac{F(-is)}{1-e^{is}} - \pi i \sum_{n= 1}^{\infty} \frac{F(- 2 n \pi i)}{(-1) 2 n \pi i}.
\end{eqnarray}
Here the principal value is taken along the real proper-time and for residues along the axis of contour in the fourth quadrant.
Finally, we obtain the effective action in the proper-time
\begin{eqnarray}
 {\cal L}_{\rm eff} &=& \frac{1}{2} \int \frac{d^3{\bf k}}{(2 \pi)^3} {\cal P} \int_{0}^{\infty} \frac{ds}{s} \Bigl[ e^{- \omega_{\rm out}T s} + e^{- \omega_{\rm in}T s} -2 e^{- \omega_{+}T s}\Bigr] \Bigl[ \frac{\cos(\frac{s}{2})}{\sin(\frac{s}{2})} \Bigr]
  \nonumber\\&&+ \frac{i}{2} \int \frac{dk}{2 \pi} \ln \Bigl[ \frac{(1 - e^{- 2 \pi \omega_{+}T})^2}{(1 - e^{- 2 \pi \omega_{\rm in}T})(1 - e^{- 2 \pi \omega_{\rm out}T})} \Bigr]. \label{con-eff}
\end{eqnarray}
Thus, the effective action (\ref{con-eff}) as a functional of $R$ and $R^{\mu \nu} R_{\mu \nu}$ gives the effective action for gravity. The vacuum persistence for bosons also holds
\begin{eqnarray}
2 {\rm Im} ({\cal L}_{\rm eff} ) = \int \frac{d^3{\bf k}}{(2 \pi)^3} \ln (1 + {\cal N}_{\bf k}), \quad {\cal N}_{\bf k} = \frac{\sinh^2 (\pi \omega_{-}T)}{\sinh (\pi \omega_{\rm in}T) \sinh (\pi \omega_{\rm out}T)}.
\end{eqnarray}

We now analyze the UV and the IR structures of the effective action (\ref{con-eff}). The
 effective gravity action is finite at $s = 0$, which can be directly shown by expanding the integrand in a power series of $s$. The leading term, which corresponds to the vacuum energy, vanishes
\begin{eqnarray}
\int \frac{d^3{\bf k}}{(2 \pi)^3}  {\cal P} \int_{0}^{\infty} \frac{ds}{s} (2 \omega_+ -\omega_{\rm out} -\omega_{\rm in})T = 0.
\end{eqnarray}
That is, the $\Gamma$-regularization automatically regulates away the vacuum energy, as shown for dS space in Ref. \cite{Kim10}. Using the power series for ${\rm cotan}(s)$,
we may express the vacuum polarization as
\begin{eqnarray}
 {\cal L}_{\rm eff} = \sum_{n = 1}^{\infty} \frac{(2n-2)!|B_{2n}|}{(2n)!T^{2n-1}} \int \frac{d^3{\bf k}}{(2 \pi)^3} \Bigl[\frac{2}{(\omega_+ )^{2n-1}} - \frac{1}{(\omega_{\rm in})^{2n-1}} -\frac{1}{(\omega_{\rm out})^{2n-1}} \Bigr]. \label{UV}
\end{eqnarray}
The next-to-leading divergent term from $n = 1$ is a change of the internal loop diagram and diverges quadratically since the momentum representation of the propagator at the coincident limit $( {\bf y} \rightarrow {\bf x})$ is
\begin{eqnarray}
D (x-y) = \int \frac{d^3{\bf k}}{(2 \pi)^3} \frac{e^{i {\bf k} \cdot ({\bf x} - {\bf y})} }{2 \omega({\bf k})}.
\end{eqnarray}
The last divergent term from $n = 2$ diverges logarithmically. These divergent terms renormalize the gravitational constant and the one-loop coupling constants, whose detailed calculations will be shown elsewhere. To get the renormalized effective action we subtract $(2/s - s/6 - s^3/360)$ from ${\rm cotan} (s/2)$.

\section{Four Dimensional de Sitter Space}

In this section we study the effective gravity action for the four-dimensional dS space with the metric
\begin{eqnarray}
ds^2 = - dt^2 + \frac{\cosh^2 (Ht)}{H^2} d \Omega_3^2. \label{dS met}
\end{eqnarray}
The Hubble constant $H = \sqrt{R/12}$ is given by the constant scalar curvature $R$ of dS space.
The massive scalar field is decomposed
by the spherical harmonics $Y_{L} (\Omega_3)$ of the Laplace operator on $S_3$,
\begin{eqnarray}
\Delta^{(3)} Y_{L} (\Omega_3) = - l(l+2) Y_{L} (\Omega_3), \quad (l = 0, 1, \cdots).
\end{eqnarray}
The multiplicity of the harmonics for each $l$ is $ (l+1)^2$. Though a comoving observer measures a thermal spectrum in the dS-invariant vacuum \cite{Bunch-Davies}, there is no actual particle production. Polyakov has recently scrutinized the selection of dS vacua in cosmology and argued that the in-/out-vacua are more physically favored than the dS-invariant vacuum \cite{Polyakov}.
For the massive scalar field $(m \geq 3H/2)$, the in- and the out-vacua are defined with respect to the asymptotically positive frequency
\begin{eqnarray}
\varphi_{L} (\mp \infty) = \frac{e^{-i \gamma H t}}{\sqrt{2 \gamma H \cosh^3 (Ht)}},
\quad \Bigl( \gamma = \sqrt{\frac{12 m^2}{R} -
\frac{9}{4}} \Bigr). \label{asym sol}
\end{eqnarray}
From the solution, modulo the spherical harmonics, with the asymptotical frequency (\ref{asym sol}),
\begin{eqnarray}
\varphi_{L} (t) &=& \frac{2^{l+\frac{3}{2}} Y_{L} \cosh^{l} (Ht) e^{(l+\frac{3}{2} - i \gamma )Ht}}{\sqrt{2 H \gamma}} F \Bigl(l+\frac{3}{2}, l+\frac{3}{2} -i \gamma, 1 - i \gamma,
- e^{2 Ht} \Bigr), \label{in sol}
\end{eqnarray}
we find the Bogoliubov coefficients as \cite{Kim10,Mottola}
\begin{eqnarray}
\mu_{L} = \frac{\Gamma (1 - i \gamma) \Gamma (- i \gamma)}{\Gamma
(l + \frac{3}{2} - i \gamma) \Gamma (1- l - \frac{3}{2} - i
\gamma)},~~
\nu_{L} = \frac{\Gamma (1 - i \gamma) \Gamma (i \gamma)}{\Gamma (l
+ \frac{3}{2}) \Gamma (1 - l -\frac{3}{2})}. \label{bog coef}
\end{eqnarray}
These coefficients satisfy the relation $|\mu_{L}|^2 - |\nu_{L}|^2 = 1$, as expected.
The number of produced scalar particles is
\begin{eqnarray}
{\cal N}_{L} = \vert \nu_{L} \vert^2 = \Bigl(\frac{1}{\sinh \pi \gamma} \Bigr)^2. \label{mean num}
\end{eqnarray}

Using the gamma function (\ref{gamma}), we find the effective action per Hubble volume and per Compton time $(W/ (mH^3/(2 \pi)^2) = {\cal L}_{\rm eff})$ in a complex plane
\begin{eqnarray}
{\cal L}_{\rm eff} (R) = i \sum_{l =
0}^{\infty} (l+1)^2
 \int_0^{\infty} \frac{ds}{s} \frac{e^{- i \gamma s}}{1 - e^{-s}} \Bigl[1
+ e^{-s} - e^{-(l + \frac{3}{2}) s} - e^{(l+\frac{1}{2})s} \Bigr].
\label{ds eff}
\end{eqnarray}
Then, the $\Gamma$-regularization (\ref{G-reg}) leads to the effective action in the proper-time integral
\begin{eqnarray}
{\cal L}_{\rm eff} (R) =  \sum_{l =
1}^{\infty} l^2 {\cal P} \int_0^{\infty} \frac{ds}{s}
\frac{e^{- \gamma s}}{\sin(\frac{s}{2})} \Bigl\{\cos (ls) - \cos(\frac{s}{2}) \Bigr\} + \frac{i}{2} \sum_{l =
1}^{\infty} l^2 \ln (1 +
{\cal N}_{L}). \label{ds eff2}
\end{eqnarray}
The effective action for dS space takes a similar form as Heisenberg-Euler/Schwinger effective action in a constant electric field. The term $\cos(s/2)$ is independent of quantum excitation $l$ and subtracts the background geometry. It is interesting to compare eq. (\ref{ds eff2}) with the effective actions from the propagator method (\ref{feyn}), which do not have the imaginary part \cite{Candelas-Raine,Dowker-Critchley,Das-Dunne}.

The momentum integral (\ref{HES eff}) or (\ref{con-eff}) is now replaced by the angular momentum sum due to the spherical symmetry of dS space. Though the proper-time integral is finite for each fixed $l$, the summation over angular momenta is infinite.
The divergent structure of the effective action requires a regularization scheme, for instance, the $\zeta$-regularization, $\sum_{l = 1}^{\infty} l^2 = \zeta (-2) = 0$, which removes the imaginary part and the subtraction of the background geometry, as shown in Ref. \cite{Kim10}. Here we use another regularization scheme for the angular momentum sum \cite{PBM}
\begin{eqnarray}
\kappa (s, n) \equiv \sum_{l =1}^{\infty} l^n \cos(l s) = \frac{1}{2 \Gamma (-n)} \int_{0}^{\infty} \frac{dt}{t^{n+1}} e^{-t} \Bigl[ \frac{e^{t} \cos(s) -1}{ \cosh (t) - \cos(s)}  \Bigr].
\end{eqnarray}
Using $\Gamma (0) =- \Gamma (-1) = 2 \Gamma (-2)$ and $\Gamma (-2) = \infty$, we have three nonvanishing terms, which in fact add up to zero
\begin{eqnarray}
\kappa (s, 2) = - \frac{1}{2} - \frac{\cos(s)}{1- \cos(s)} + \frac{1+ \cos(s)}{2(1-\cos(s))} = 0.
\end{eqnarray}
The $\kappa$-regularization is equivalent to the $\zeta$-regularization after expanding $\cos(ls)$ in power series of $s$ and putting $\zeta (- 2n) = 0$.
Finally, the vacuum polarization takes the form
\begin{eqnarray}
{\cal L}_{\rm eff} (R) =  - \kappa (0,2) {\cal P} \int_0^{\infty} \frac{ds}{s} e^{- \gamma s}\Bigl[\frac{\cos(\frac{s}{2})}{\sin(\frac{s}{2})} \Bigr]. \label{ds ren}
\end{eqnarray}
As $\kappa (0,2) = \zeta (-2) =0$, the effective action vanishes for dS space in $D=4$, implying no quantum hair
and confirming the result of Ref. \cite{Kim10}.

\section{Conclusion}

We have explored the vacuum structure of a conformally, asymptotically flat spacetime and a de Sitter spacetime in $D=4$. The gamma function or the $\Gamma$-regularization has been used to provide effective actions not only for QED but also for gravity. The key idea is to regularize the logarithm of gamma functions for the effective action that is expressed in terms of the Bogoliubov coefficient. Remarkably, the effective action for gravity takes a similar form as the Heisenberg-Euler and Schwinger one for QED. We have analyzed the UV and the IR structures of the effective action for gravity.
The divergent terms are closely related with renormalization of the vacuum energy, the gravitational constant and the one-loop coupling constants. One surprising result is that the de Sitter space in $D=4$ does not have any quantum hair due the maximal symmetry of spacetime \cite{Kim10}.

\section*{Acknowledgments}

The author would like to thank W-Y.~Pauchy Hwang for warm hospitality at National Taiwan University and also thank Remo Ruffini for warm hospitality at ICRANet and supporting the participation of the second Galileo-Xu Guangqi Meeting, Italy, July 12-17, 2010. The visit to the Institute of Astrophysics, National Taiwan University, was supported by National Science Council Grant NSC100-2811-M-002-012.
This work was supported by the National Research Foundation (NRF) Grant funded by the
Korean Government (MEST)(2010-0016-422).


\begin{thebibliography}{99}

\bibitem{Heisenberg-Euler} W.~Heisenberg and H.~Euler, Z. Phys. {\bf 98}, 714
(1936).

\bibitem{Schwinger} J.~Schwinger, Phys. Rev. {\bf 82}, 664 (1951).

\bibitem{Ringwald} A. Ringwald, Phys. Lett. B {\bf 510}, 107 (2001) [arXiv:hep-ph/0103185
].

\bibitem{Schwinger-var} J.~Schwinger, Proc. Natl. Acad. Sci. (U.S.A.) {\bf 37}, 452
(1951); Proc. Natl. Acad. Sci. (U.S.A.) {\bf 37}, 455
(1951).

\bibitem{Nikishov} A.~I.~Nikishov, Nucl. Phys. B {\bf 21}, 346 (1970); Phys. Atom. Nucl. {\bf 67}, 1478 (2004) [arXiv:hep-th/0304174].

\bibitem{DeWitt} B.~S.~DeWitt, Phys. Rep. {\bf 19}, 295 (1975).

\bibitem{Birrel-Davies} N.~D.~Birrell and P.~C.~W.~Davies, {\it Quantum Fields in Curved
Space} (Cambridge University Press, Cambridge, UK, 1984).

\bibitem{KLY08} S.~P.~Kim, H.~K.~Lee, and Y.~Yoon, Phys. Rev. D {\bf 78},
105013 (2008) [arXiv:hep-th/0807.2696]; S.~P.~Kim and H.~K.~Lee, J. Korean
Phys. Soc. {\bf 54}, 2605 (2009) [arXiv:hep-th/0806.2496].

\bibitem{AHN} J.~Ambj{o}rn, R.~J.~Hughes, and N.~K.~Nielsen, Ann. Phys. {\bf 150}, 92 (1983).

\bibitem{Nikishov03} A.~I.~Nikishov, JETP {\bf 96}, 180 (2003) [arXiv:hep-th/0207085].

\bibitem{KLY10} S.~P.~Kim, H.~K.~Lee, and Y.~Yoon, Phys. Rev. D {\bf 82}, 025015 (2010) [arXiv:hep-th/0910.3363].

\bibitem{Kim10} S.~P.~Kim, ``Vacuum Structure of  de Sitter Space,'' [arXiv:hep-th/1008.0577].

\bibitem{Polyakov} A.~M.~Polyakov, Nucl. Phys. B {\bf 797}, 199 (2008) [arXiv:hep-th/0709.2899]; Nucl. Phys. B {\bf 834}, 316 (2010) [arXiv:hep-th/0912.5503].

\bibitem{Dunne-rev} G.~V.~Dunne, in {\it From Fields to Strings: Circumnavigating
Theoretical Physics}, edited by M.~Shifman, A.~Vainshtein, and
J.~Wheater (World Scientific, Singapore, 2005), Vol. I, pp. 445-522
[arXiv:hep-th/0406216].

\bibitem{Dunne-Hall99} G.~Dunne and T.~Hall, Phys. Rev. D {\bf 60},
065002 (1999) [arXiv:hep-th/9902064].

\bibitem{GGT} S.~P.~Gavrilov, D.~M.~Gitman, and J.~L.~Tomazelli, Nucl. Phys. B {\bf 795}, 645 (2008) [arXiv:hep-th/0612064].

\bibitem{Hwang-Kim} W-Y.~P.~Hwang and S.~P.~Kim, Phys. Rev. D {\bf
80}, 065004 (2009) [arXiv:0906.3813].

\bibitem{Gamma} I.~S.~Gradshteyn and I.~M.~Ryzhik, {\it Table of Integrals, Series,
and Products} (Academic Press, San Diego, 1994), p. 8.341-2.

\bibitem{Bunch-Davies} T.~S.~Bunch and P.~C.~W.~Davies, Proc. R. Soc. London A {\bf 360}, 117 (1978).

\bibitem{Mottola} E.~Mottola, Phys. Rev. D {\bf 31}, 754 (1985).

\bibitem{Candelas-Raine} P.~Candelas and D.~J.~Raine,
Phys. Rev. D {\bf 12}, 965 (1975); Phys. Rev. D {\bf 15}, 1494 (1977).

\bibitem{Dowker-Critchley}J.~S.~Dowker and R.~Critchley,
Phys. Rev. D {\bf 13}, 224 (1976); Phys. Rev. D {\bf 13}, 3224 (1976).

\bibitem{Das-Dunne} A.~Das and G.~V.~Dunne, Phys. Rev. D {\bf 74},
044029 (2006) [arXiv:hep-th/0607168].

\bibitem{PBM} A.~P.~Prudnikov, Yu.~A.~Brychkov, and O.~I.~Marichev,
{\it Integrals and Series} (Gordon and Breach Science Publishers,
The Netherlands, 1998) Vol. 1, p 725.

\end{thebibliography}
\end{document}